\documentclass[12pt]{iopart}
\pdfoutput=1
\usepackage{iopams}
\usepackage{graphicx}
\usepackage{cite} 

\begin{document}

\title{Spectral self-action of THz emission from ionizing two-color laser pulses in gases}

\author{Eduardo Cabrera-Granado$^1$, 
Yxing Chen$^2$,
Ihar Babushkin$^3$,
Luc Berg\'e$^4$,
and Stefan Skupin$^{5,6,7}$}

\address{$^1$Facultad de \'Optica y Optometr\'ia, Universidad Complutense de Madrid, 28037,
Madrid, Spain}
\address{$^2$Laboratory for fundamental BioPhotonics, Ecole polytechnique f\'ed\'erale de Lausanne, Switzerland}
\address{$^3$Humboldt University, Institute of Mathematics, 12489 Berlin, Germany}
\address{$^4$CEA-DAM, DIF, F-91297 Arpajon, France}
\address{$^5$Max Planck Institute for the Physics of Complex Systems, 01187 Dresden,
Germany}
\address{$^6$Institute of Condensed Matter Theory and Solid State Optics, Abbe Center of Photonics, Friedrich-Schiller-Universit\"at Jena, 07743 Jena, Germany}
\address{$^7$Univ.~Bordeaux - CNRS - CEA, Centre Lasers Intenses et Applications, UMR 5107, 33405 Talence, France}

\ead{ecabrera@ucm.es}

\begin{abstract}
The spectrum of terahertz (THz) emission in gases via ionizing two-color femtosecond pulses is analyzed by means of a semi-analytic model and finite-difference-time-domain simulations in 1D and 2D geometries.
We show that produced THz signals interact with free
electron trajectories and thus influence significantly further THz generation upon propagation, i.e., make the process inherently nonlocal.
This self-action plays a key role in the observed strong spectral broadening 
of the generated THz field.
Diffraction limits the achievable THz bandwidth
by efficiently depleting the low frequency amplitudes in the propagating field.

\end{abstract}

\submitto{\NJP}

\maketitle

\section{Introduction}
\label{sec:intro}

Research on intense terahertz (THz) electromagnetic sources has received an increasing attention owing
to numerous applications, for example, in time-domain spectroscopy, biomedical imaging or security screening~\cite{Tonouchi:np:1:97}. Among the various techniques employed
to generate THz radiation, focusing intense two-color femtosecond pulses in air or noble gases provides interesting features like absence of material damage, large generated bandwidth (up to $\sim100$~THz) and high amplitudes of the emitted THz pulses ($> 100$~MV/m)~\cite{Kim:ieeejqe:48:797}. 
First reported by Cook {\it et al.}~\cite{Cook:ol:25:1210}, 
THz emission from intense two-color pulses was initially attributed to optical rectification via third-order nonlinearity. However, it was shown later that the plasma built-up
by tunneling photoionization is necessary to explain the high amplitudes of the THz field \cite{Kress:ol:29:1120,Roskos:lpr:1:349,Kim:np:2:605}, and a quasi-dc plasma current generated
by the temporally asymmetric two-color field is responsible for THz emission~\cite{berge:prl:110:073901,Borodin:ol:38:1906}. 
Plasma oscillations leading to strong THz radiation were also reported for single color pump pulses with few-cycle duration or at higher intensities~\cite{chen:pre:78:046406,Wang:ol:36:2608}.

Apart from energy scaling \cite{oh:njp:15:075002} and polarization control \cite{wen:prl:103:023902}, tailoring the shape of the broadband radiated THz pulse is one of the standing goals with respect to applications. In the case of two-color filaments in air it was already demonstrated experimentally that the geometry of the plasma channels and the initial carrier envelope phase of the laser pulse can be used to control the THz waveform~\cite{Manceau:ol:34:2165,Bai:prl:108:255004}. 
It was also recently suggested to control THz generation in gases by more involved spectral engineering of the IR pump pulse, i.e., by modifying the temporal positions of
the electric field maxima resp.\ ionization events \cite{Babushkin:njp:13:123029}. 

One of the main challenges on the route towards THz spectral control is to understand the influence of the complicated nonlinear propagation dynamics of the electromagnetic radiation.
It is known that ionizing femtosecond laser pulses undergo strong spatiotemporal modifications during propagation, and that these propagation effects have a tremendous impact on the emitted THz fields~\cite{Babushkin:prl:105:053903}. Moreover, shortly after the onset of THz generation the gas atoms or molecules and ionized electrons are exposed to the co-propagating low-frequency field as well. In particular because asymmetrically ionized gases were already successfully used for remote detection of THz fields through coherent manipulation of the ionized electron drift velocity and subsequent collision-induced fluorescence emission~\cite{Liu:np:4:627}, we can suspect to find a self-action mechanism of already generated THz radiation on the THz generation itself.

\section{Model}
\label{sec:model}

\begin{figure}
\centerline{\includegraphics[width=\columnwidth]{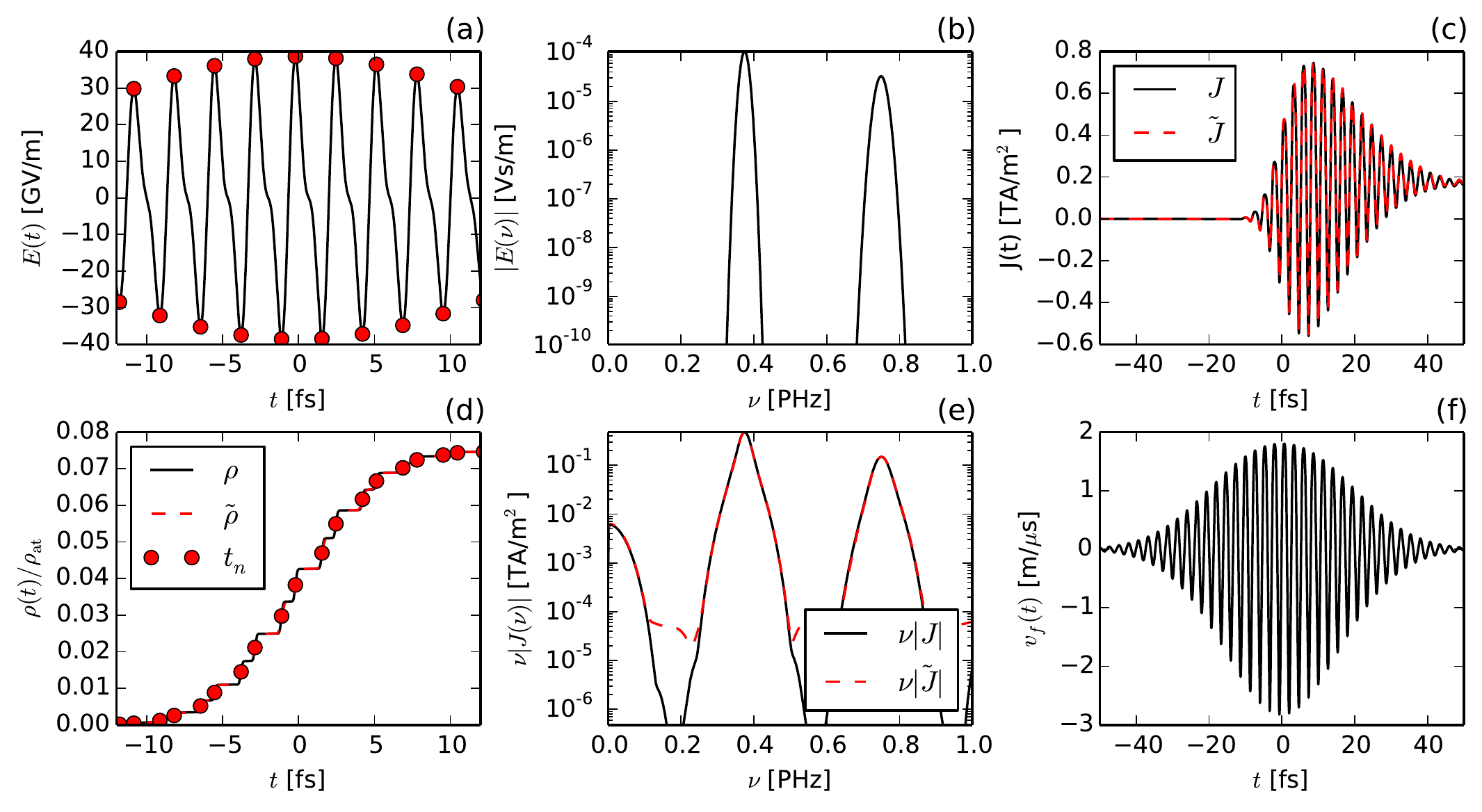}}
\caption{(a) Exemplary two-color electric field $E(t)$ [see Eq.~(\ref{eq:initialfield}) and text below for details]; (b) corresponding spectrum; (c) current density $J(t)$ and (d) plasma density $\rho(t)$ given by Eqs.~(\ref{eq:J}) and (\ref{eq:rho}). (e) The resulting spectrum of the secondary electric field $E_J \propto \nu J$ [see Eq.~(\ref{eq:ej})], and (f) the free electron velocity $v_f$ [see Eq.~(\ref{eq:v_f})].
Results of our semi-analytic model Eqs.~(\ref{eq:rho_approx}) resp.\ (\ref{eq:current_approx}),  $\tilde{\rho}(t)$ and $\tilde{J}(t)$, are marked as red dashed lines in (c),(d), and (e), showing excellent agreement. Red dots in (a) resp.\ (d) indicate
the temporal positions t$_n$ of the ionization events.}
\label{fig0}
\end{figure}

In the present work we will shed light onto the pulse propagation effects in the plasma, and the interaction of the generated terahertz field with the ionized medium. Our starting point is a semi-analytic model developed in~\cite{Babushkin:njp:13:123029} based on the local current (LC) approximation, i.e., considering a small volume of gas irradiated by the ionizing field. 
Let us assume that the free electron density is governed by 
\begin{equation}\label{eq:rho}
\partial_t \rho(t) = W_{ST}(E)[\rho_{at} - \rho(t)],
\end{equation}
where $W_{ST}(E)$ is a field-dependent tunneling ionization rate \cite{Roskos:lpr:1:349}, leading to a stepwise increase of $\rho(t)$ in time [see Fig.~\ref{fig0}(d)]. The $n$-th ionization event with amplitude $\delta \rho_n$ and temporal shape $H_n(t)$ corresponds to a maximum of the incoming field at time $t_n$. Because all ionization events share a similar shape~\footnote{Provided that pump pulses are multi-cylce, and $\rho(t)\ll\rho_{at}$.}, we can simplify $H_n(t) \simeq H(t - t_n) = \left\{1 + \mathrm{erf}[(t-t_n) /\tau]\right\}/2$, with characteristic temporal width $\tau = 0.2$~fs~\cite{Babushkin:njp:13:123029}. The events are well separated in time, so we can give a semi-analytic expression for $\rho(t)$ by summing up all
contributions
\begin{equation}\label{eq:rho_approx}
\rho(t) \approx \tilde{\rho}(t) = \sum_n \delta \rho_n H(t-t_n).
\end{equation}
For a given electric field amplitude $E(t)$, the $\delta \rho_n$ and $t_n$ can be extracted from the numerical solution of Eq.~(\ref{eq:rho}).

If we assume zero velocity for 
newly born electrons and neglect ponderomotive forces, the equation for the plasma current density reads
\begin{equation}\label{eq:J}
 \frac{dJ(t)}{dt} + \gamma J(t) =  \frac{q^2}{m} \rho(t) E(t).
\label{eq:current}
\end{equation}
Here, $\gamma=7.7$~ps$^{-1}$ is a phenomenological electron-ion collision rate, and $q,m$ represent electron charge and mass, respectively. 
Finally, plugging the above approximation $\tilde{\rho}(t)$ [Eq.~(\ref{eq:rho_approx})] for the plasma density $\rho(t)$ into Eq.~(\ref{eq:current}) yields a semi-analytic expression for the current 
\begin{equation}\label{eq:current_approx}
J(t) \approx \tilde{J}(t)=\sum_n q \delta \rho_n H(t-t_n) \left[v_f(t) -e^{\gamma(t_n-t)} v_f(t_n)\right].
\end{equation}
Here, the expression
\begin{equation}\label{eq:v_f}
v_f(t) = \frac{q}{m} \int_{-\infty}^t E(\tau) e^{\gamma(\tau-t)} \mathrm{d}\tau
\end{equation}
can be interpreted as the free electron velocity. 
Figure~\ref{fig0} confirms excellent agreement between numerical evaluation of Eqs.~(\ref{eq:rho}), (\ref{eq:current}) and our semi-analytic expression Eqs.~(\ref{eq:rho_approx}), (\ref{eq:current_approx}) for both plasma and current density. In fact, throughout the whole analysis presented in this paper we found that the approximations $\rho(t)\approx\tilde{\rho}(t)$ and $J(t)\approx\tilde{J}(t)$ are always close to equality.

The emitted secondary electric field due to the plasma current can be calculated in 
frequency domain as 
\begin{equation} \label{eq:ej}
E_J(\omega) =g \omega J(\omega), 
\end{equation}
where $g$ is a constant~\cite{Jefimenko:EM:66}.
In Fig.~\ref{fig0}(e) we show the spectrum of the secondary radiation according to Eq.~(\ref{eq:ej}) for a representative (linearly polarized) two-color pump field 
\begin{equation} \label{eq:initialfield}
 E(t) = E_0 \left[ e^{-\frac{t^2}{t_p^2}}\cos\left(\omega t\right) + r e^{-\frac{t^2}{2 t_p^2}} \cos\left(2 \omega t + \phi \right)\right],
\end{equation}
where $E_0 = 31$~GV/m, $t_p = 24$ fs, $\omega = 2 \pi \nu$ with $\nu = 375$~THz, and the ratio $r=0.44$ between the fundamental and second-harmonic. The relative phase between both fields has been set to $\phi=\pi/2$, to ensure optimum conditions for THz generation~\cite{Roskos:lpr:1:349}.
Throughout this paper all quantities, including spectra, are expressed in physical units.

\section{THz spectral self-action}
\label{sec:selfaction}

\begin{figure}
\centerline{\includegraphics[width=\columnwidth]{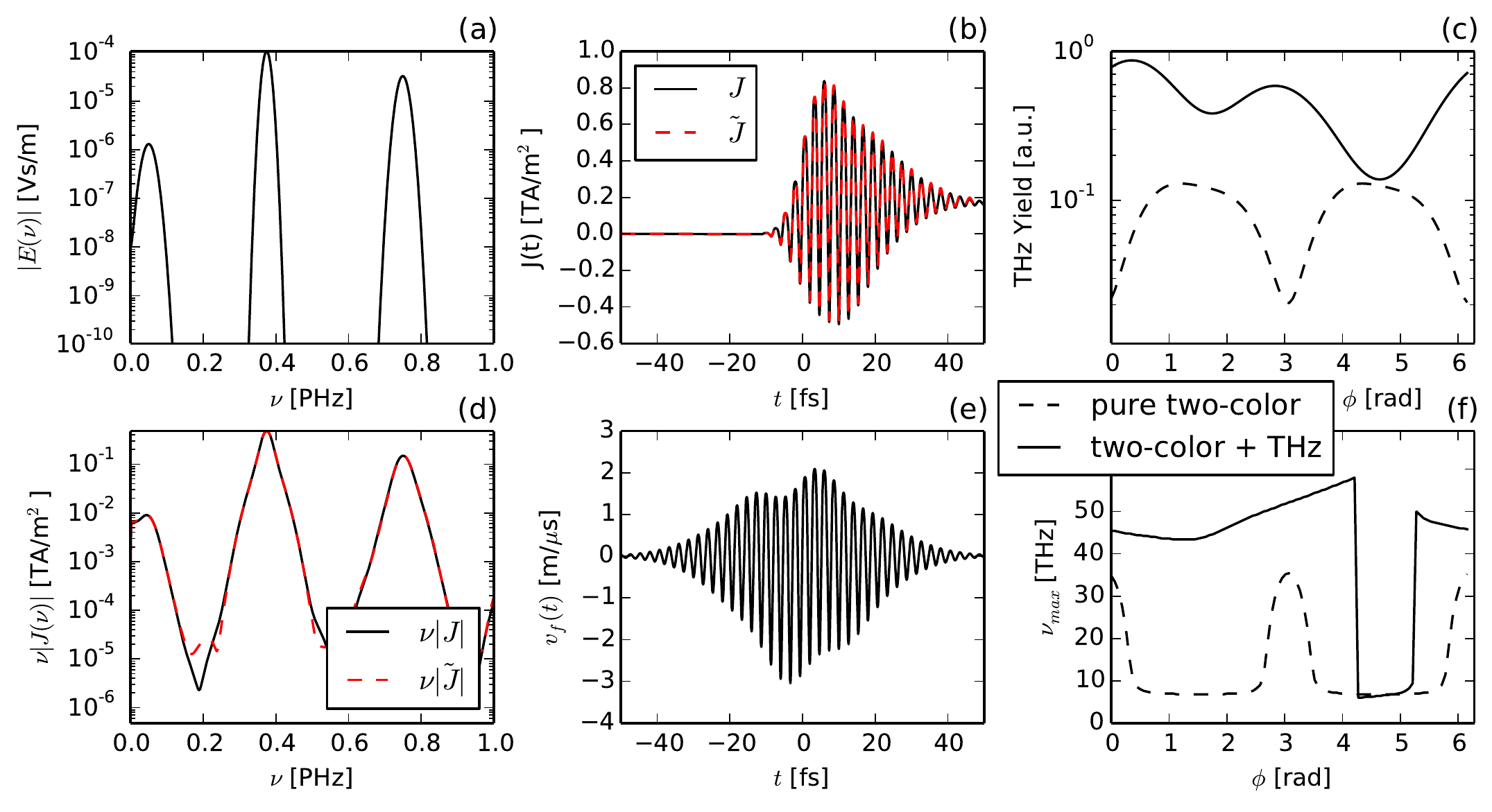}}
\caption{(a) Spectrum of the exemplary pump field with THz component centered at 50~THz (see text for details); (b) corresponding current density $J(t)$; (c) yield of secondary radiation below 100~THz for the pump field in (a) (solid curve) and the pure two-color field of Fig.~\ref{fig0} versus phase angle $\phi$.
(d) Spectrum of the secondary electric field $E_J \propto \nu J$ and (e) free electron velocity $v_f$ for the pump field in (a).
(f) Position of the maximum spectral density of secondary radiation below 100~THz, same line coding as in (c). Results of our semi-analytic model are marked as red dashed lines in (b) and (d), showing again excellent agreement.}
\label{fig1bis}
\end{figure}

It is a reasonable assumption that an additional small low frequency field will co-propagate with the pump pulse shortly after the onset of THz generation. Thus, let us now investigate the impact of such field on the secondary radiation spectrum. 
To this end, we add a third component centered at 50~THz, $\sim 15$~fs duration
and with only 2~\% the amplitude of the fundamental IR frequency to the two-color field Eq.~(\ref{eq:initialfield}) [see Fig.~\ref{fig1bis}(a)]. 
As can be seen in Fig.~\ref{fig1bis}(d), the low-frequency spectral shape of the secondary radiation changes noticeably compared to Fig.~\ref{fig0}(e), a peak around the frequency of the new pump component now dominates the low frequency range. This simple example already indicates that generated THz fields have an important impact on the subsequent THz generation process, and thus produce a self-action. In other words, the THz generation process from ionizing two-color pulses is nonlinear, and the nonlinearity is significantly nonlocal in propagation direction of the pump laser.

Before doing any further analysis, we want to investigate the influence of the phase angle $\phi$ between fundamental and SH field. The secondary radiation spectra shown in Figs.~\ref{fig0}(e) and \ref{fig1bis}(d) are obtained for $\phi=\pi/2$. It is well known that for a pure two-color driving field this value ensures maximum THz yield $\propto \int_0^{100~\textrm{\scriptsize{THz}}} \nu^2 |E(\nu)|^2 d\nu$, as recalled by the dashed line in Fig.~\ref{fig1bis}(c). With the small THz component present in the driving field, this THz yield increases considerably for almost all values of $\phi$ [see solid curve in Fig.~\ref{fig1bis}(c)], and the frequency of the maximum spectral density increases as well [see Fig.~\ref{fig1bis}(f)]. 
The jump in Fig.~\ref{fig1bis}(f) around $\phi=3\pi/2$ is linked to the exceptionally low THz yield in this parameter range [cf.\ Fig.~\ref{fig1bis}(c)].
Thus, we can state that the additional THz pump field component, even though with small amplitude, dominates the low frequency spectral shape of the secondary emission for almost all values of $\phi$. We also found that the phase angle of the THz pump component itself is of minor influence (not shown), which is probably due to its much longer cycle duration.

\begin{figure}
\centerline{\includegraphics[width=0.66\columnwidth]{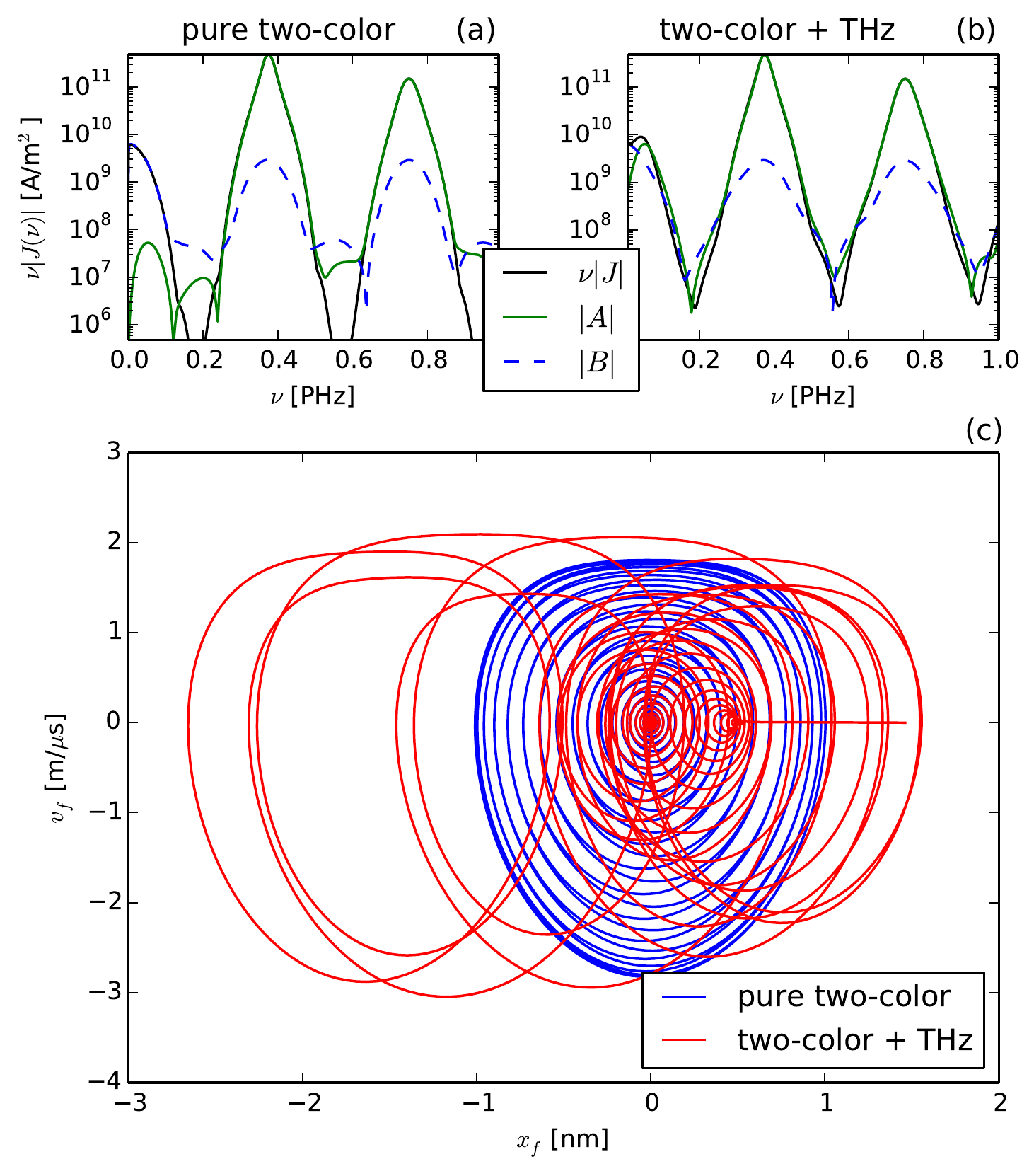}}
\caption{Spectrum of secondary radiation $E_J(\nu) \propto \nu J(\nu)$ (black solid line) for (a) a pure two-color pump pulse (cf.~Fig.~\ref{fig0}) and (b) with an additional low
amplitude component centered around 50~THz (cf.~Fig.~\ref{fig1bis}). Our semi-analytic model Eqs.~(\ref{eq:A}), (\ref{eq:B}) allows a decomposition into $A(\nu)$ (green solid lines) and $B(\nu)$ (blue dashed lines). The term $A(\nu)$, which contains the impact of the electric field on the plasma current via the free electron velocity $v_f(t)$, is clearly responsible for the modification of the secondary radiation in THz spectral range. In (c) the phase-space representation of the motion of a free electron born at $t=-\infty$ is shown for both pump field configuations. The dramatic impact of the THz component in the electron trajectory is clearly visible.} 
\label{fig1}
\end{figure}

We can use the semi-analytic approximate expression for the plasma current $\tilde{J}(t)$, Eq.~(\ref{eq:current_approx}), to get further insight into the THz spectral self-action mechanism. To this end, we rewrite the secondary field as the sum of two contributions,  
\begin{equation}
 E_J(\omega) = g \omega {J}(\omega) \approx g \omega \tilde{J}(\omega) =  g\left[A(\omega) - B(\omega)\right],
 \label{eq:fieldj}
\end{equation}
where
\begin{eqnarray}
A(\omega) & = q\omega\mathrm{FT}\left[v_f(t) \sum_n \delta \rho_n H(t-t_n) \right], \label{eq:A}\\
B(\omega) & = q\omega\mathrm{FT}\left[\sum_n \delta \rho_n H(t-t_n) e^{\gamma(t_n-t)} v_f(t_n)\right]. \label{eq:B}
\end{eqnarray}
Here, FT[~] denotes the Fourier transform. Interestingly, for a two-color pump field 
without any low-frequency components in optimum configuration ($\phi=\pi/2$, cf.\ Fig.~\ref{fig0}) the secondary radiation $E_J(\nu)$ below 100~THz is determined solely by $B(\nu)$~\cite{Babushkin:njp:13:123029}. This is confirmed by Fig.~\ref{fig1}(a), where the secondary radiation $E_J(\nu) \propto \nu J(\nu)$ is separated into $A(\nu)$ and $B(\nu)$ according to Eqs.~(\ref{eq:A}), (\ref{eq:B}). Once low-frequency components are present, $A(\nu)$ starts to contribute as well through the free electron velocity $v_f(t)$. Figure~\ref{fig1}(b) illustrates the impact of $A(\nu)$ and $B(\nu)$ for the pump field configuration of Fig.~\ref{fig1bis} and reveals the THz spectral self-action mechanism:
The term $A(\nu)$ describes the impact of the electric field on the plasma current, because $A(\nu)$ contains $v_f(t)$~\footnote{In contrast to $B(\nu)$, which contains $v_f(t_n)$ only.}. In fact, the change in the free electron velocity $v_f(t_n)$ is clearly visible when comparing Fig.~\ref{fig0}(f) and Fig.~\ref{fig1bis}(e). As we can clearly observe from Fig.~\ref{fig1}(c), the low frequency component of $E(t)$ resp.\ $v_f(t)$ can significantly alter the free electron trajectory and thus the secondary radiation spectrum. 
From the mathematical structure of Eq.~(\ref{eq:A}) we can infer that 
$A(\nu)$ yields a similar spectrum as the pump field $E(\nu)$ but broader
due to the convolution of $\mathrm{FT}[v_f(t)]$ with $\mathrm{FT}[H(t)]$ in Fourier domain, which is confirmed by Fig.~\ref{fig1}. Because the secondary radiation is produced continuously added to the pump field, a spectral self-action occurs and we expect a THz spectral broadening upon propagation.

\section{Simulations}
\label{sec:simulations}

\begin{figure}
 \centerline{\includegraphics[width=\columnwidth]{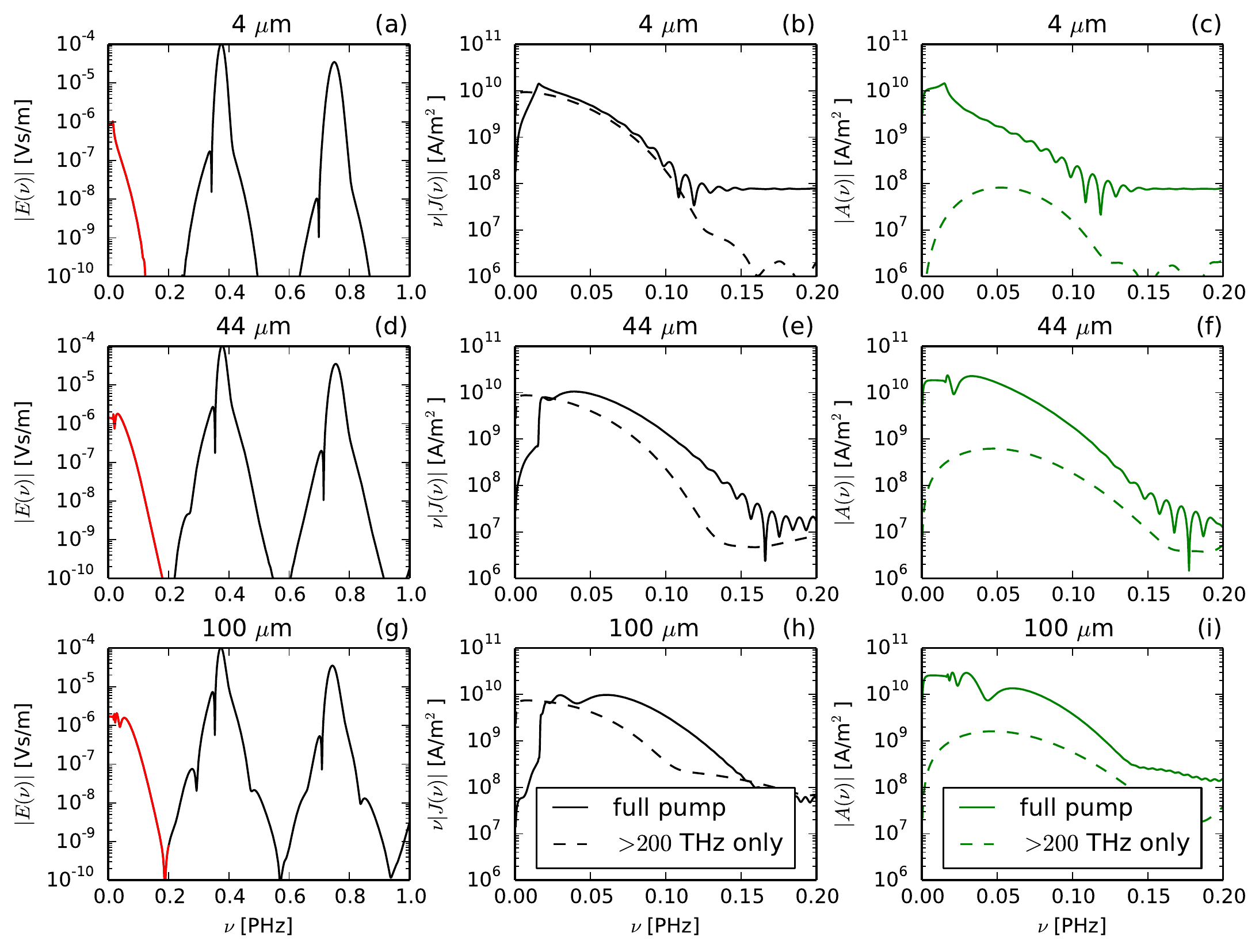}}
\caption{(a,d,g) Recorded electric field spectrum from 1D-FDTD simulations at indicated propagation distances, with low frequency part marked in red. (b,e,h) Corresponding local secondary emission $E_J(\nu) \propto \nu J(\nu)$ in the range below 200~THz (solid black lines). The dashed lines show the same quantity computed from a modified pump field with low frequency part ($<200$~THz) suppressed.
(c,f,i) The contribution $A(\nu)$ to the local secondary emission according to Eq.~(\ref{eq:A}) computed with (solid green lines) and without (dashed green lines) low frequency part of the pump.}
\label{fig2}
\end{figure}

To confirm our previous hypothesis of THz spectral self-action, we present several numerical simulations. In a first attempt, we propagate the two-color field Eq.~(\ref{eq:initialfield})~\footnote{In all simulations no THz component is present at $z=0$.} over 400~$\mu$m in argon gas by means of the 1D-finite-difference-time-domain (1D-FDTD) algorithm~\cite{Taflove:CE:95}. 
Nonlinear generalization of the FDTD algorithm offers the possibility to simulate Maxwell's equations without further approximations~\cite{etrich:pra:84:021808}.
Linear dispersion of argon is included via the refractive index $n(\omega)$ given in~\cite{Dalgarno:procrsoca:259:424}. The plasma density $\rho(t)$ obeys Eq.~(\ref{eq:rho}),
and the resulting plasma current $J(t)$ is accounted for via Eq.~(\ref{eq:current}). 
Figure~\ref{fig2}(a,d,g) shows the spectrum of the propagated field $E(\nu)$ at three different distances. 
We can clearly see that the low frequency spectrum (red part of the curve) broadens up to frequencies well above 100~THz during propagation in the medium, in agreement with our previous expectations. 
Solid black lines in Fig.~\ref{fig2}(b,e,h) show the corresponding local secondary emission $E_J(\nu) \propto \nu J(\nu)$.
Interestingly, the maximum of the low frequency secondary emission spectrum shifts toward larger frequencies with increasing propagation distance, an effect already reported in~\cite{Babushkin:prl:105:053903,babushkin:oe:18:9658}.

In order to further corroborate the THz spectral feedback mechanism, hypothetical local secondary emission spectra computed from a modified pump field with low frequency part ($<200$~THz) suppressed are plotted in Fig.~\ref{fig2}(b,e,h) for comparison (dashed lines).
Obviously, local spectra generated by these artificially modified pump fields are less broad than the original spectra, in particular towards larger propagation distances where strong THz spectral broadening takes place. 
By using the semi-analytic model developed above, we can decompose the local secondary emission $\nu J(\nu)$ into the contributions of $A(\nu)$ and $B(\nu)$. As predicted above and confirmed by Fig.~\ref{fig2}(c,f,i), the low frequency part of $A(\nu)$ is determined by the low frequency part of the driving field $E$, and thus responsible for the THz spectral feedback mechanism. In contrast, we report minor changes only in $B(\nu)$ when the low frequency part of the pump field is suppressed (not shown).

Thus, in the present propagation regime with relatively narrow IR pump spectra, the THz spectral feedback mechanism is a key player with
strong impact on the emitted secondary radiation. At larger propagation distances, spectral broadening and shifting of the two-color pump pulse itself in IR and visible domain may become the dominant effects in shaping the local secondary emission THz spectra (not shown). In fact, it is important to keep in mind that the THz spectral feedback mechanism reported here is an additional effect, and other propagation effects modifying the THz 
spectra are present as well~\cite{Babushkin:prl:105:053903,berge:prl:110:073901}.

\begin{figure}
\centerline{\includegraphics[width=\columnwidth]{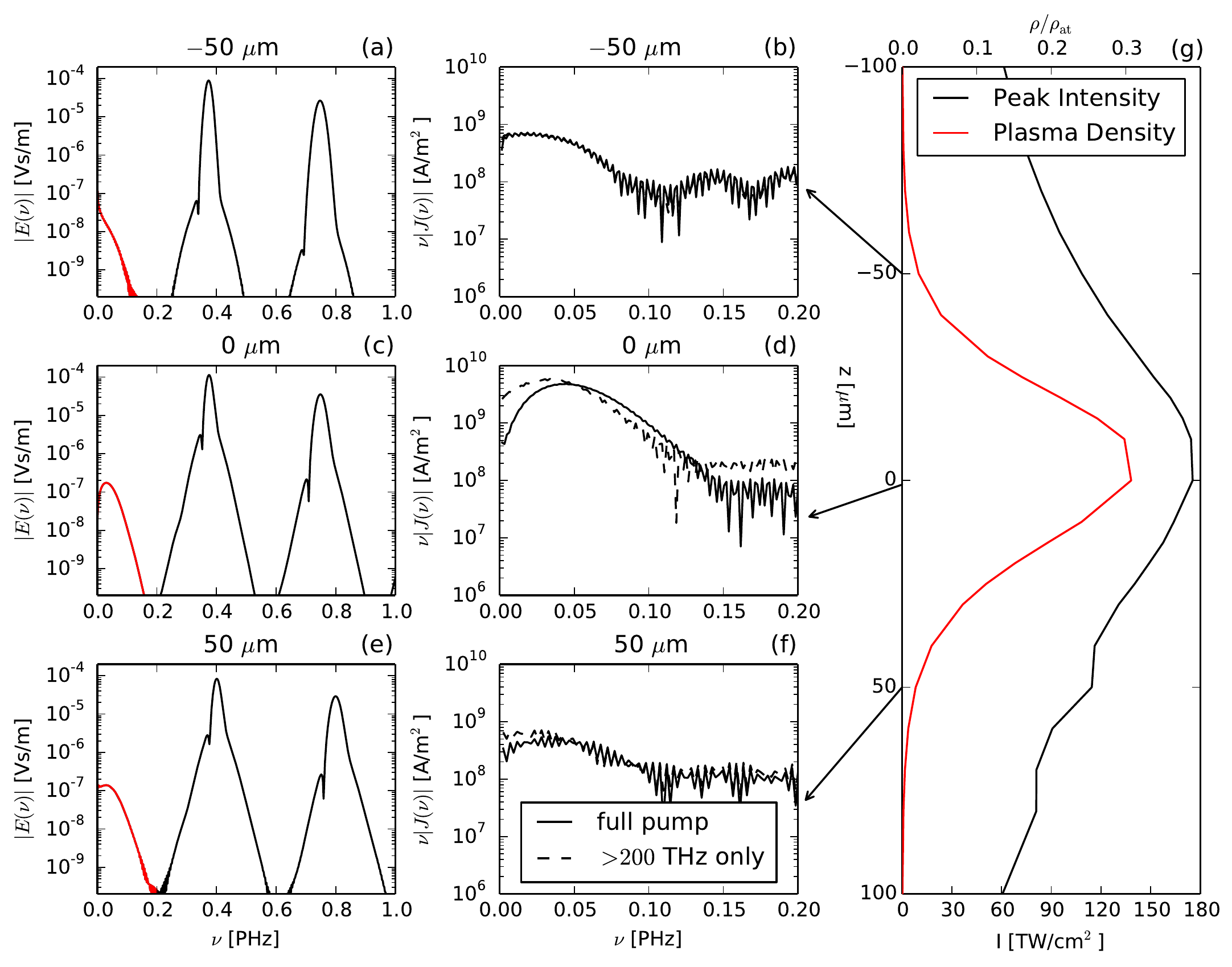}}
\caption{(a,c,e) Recorded on-axis electric field spectrum from 2D-FDTD simulations at indicated propagation distances, with low frequency part marked in red. (b,d,f) Corresponding on-axis local secondary emission $E_J(\nu) \propto \nu J(\nu)$ in the range below 200~THz (solid black lines). The dashed lines show the same quantity computed from a modified pump field with low frequency part ($<200$~THz) suppressed.
(g) Peak intensity and plasma density versus propagation distance $z$, with $z=0$ denoting the focal point.}
\label{fig3}
\end{figure}

In the 1D diffraction-free geometry investigated so far, nonlinear effects are more pronounced than in 2D or even 3D geometries with strongly varying pump intensities due to focusing and defocusing of the pump.
2D-FDTD simulation results employing a focused ($f=330~\mu$m) beam with initial width $w_0=30.2~\mu$m are shown in Fig.~\ref{fig3}. The initial fundamental amplitude is fixed to 11.5~GV/m, in order to reach 47~GV/m peak amplitude at focus. Other parameters are kept as in the 1D simulations [see Eq.~(\ref{eq:initialfield})].
The interaction of the generated THz field with the newly born free electrons as the field propagates through the medium is expected to be weaker than in 1D configuration, because the low frequency part of the field strongly diffracts and leaves the plasma channel. 
Nevertheless, as the field propagates the maximum of the secondary radiation spectrum in THz range shifts towards higher frequencies until the focal point is reached [see Fig.~\ref{fig3}(a,c,e)].
Dashed curves in Fig.~\ref{fig3}(b,d,f) show on-axis local secondary emission spectra $E_J(\nu) \propto \nu J(\nu)$ computed from pump fields with low frequency range $<200$~THz suppressed. Local spectra generated near the focus by these artificially modified pump fields are shifted by more than 10~THz towards lower frequencies compared to the original local spectra plotted in black solid lines. We note that in simulations reaching higher peak electric field amplitudes $>50$~GV/m at focus the THz self-action can be more pronounced, however, our model accounting for single ionization only becomes questionable in this high-intensity regime. In contrast, simulations with reduced fundamental peak amplitude reaching only 37~GV/m at focus (not shown) feature no THz spectral self-action at all, simply because the on-axis THz field amplitude remains one order of magnitude lower.

\begin{figure}
\centerline{\includegraphics[width=0.75\columnwidth]{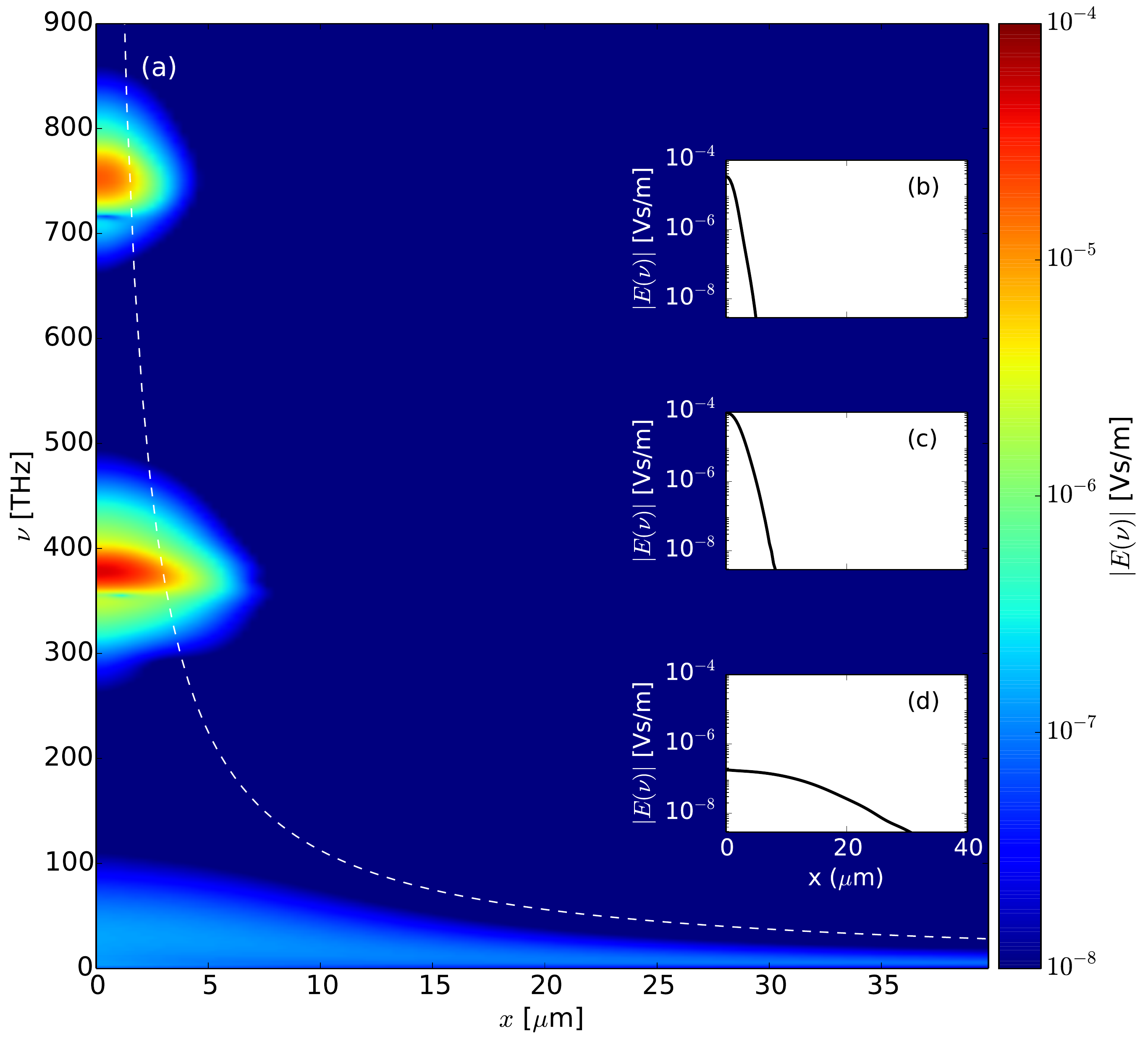}}
\caption{(a) Spatially resolved electric field spectrum (log scale) of the propagating beam in 2D-FDTD simulations obtained 30~$\mu$m after the geometrical focus.
Because of symmetry we show positive transverse $x$-axis only.
Line-outs of spatial profiles versus coordinate $x$ for frequencies $\nu = 375$ THz, $\nu = 750$ THz and $\nu = 30$ THz are shown in panels (b), (c) and (d), respectively. The white dashed line in (a) serves as an eye-guide and shows the expected beam width $w(\nu)\propto 1/\nu$ for a perfect Gaussian beam at focus for all frequencies.  
Beam parameters are the same as in Fig.~\ref{fig3}.}
\label{fig4}
\end{figure}

In general, THz self-action in 2D geometry is less pronounced than in the former 1D case. Moreover, from the focal point onward, Fig.~\ref{fig3}(d,f) shows that the maximum of the secondary field spectrum remains fixed around $\nu \sim 50$~THz. These features can be readily explained by the strong defocusing of
the THz field beyond focus, which prevents the low frequency field to drive the free electrons produced in the plasma channel, and therefore arrests the THz spectral self-action mechanism. 
In Fig.~\ref{fig4} the action of diffraction on the different spectral components of the electric field $30~\mu$m after the geometrical focus is visualized. One can clearly see that diffraction delocalizes the THz radiation much stronger than the IR and visible light. Thus, diffraction is a key player in determining the final bandwidth of the secondary emission, because it rapidly decreases the on-axis THz field strength.

\section{Conclusions}
\label{sec:conclusions}

In conclusion, we have investigated THz emission via ionizing two-color femtosecond pulses and revealed a 
THz spectral self-action mechanism. This mechanism stems from the interaction of the already generated
THz field with the subsequent free electron dynamics in the plasma channel and 
leads to strong THz spectral broadening. Direct FDTD simulations in 1D and 2D geometries support 
our semi-analytic model and show that diffraction is a key player in determining the final bandwidth of the secondary emission by limiting the interaction length of the low-frequency field with the
plasma. We believe that our findings may have implications beyond the generation of broadband THz radiation, namely, on the interpretation of recent experiments on high harmonic generation with two-color pulses~\cite{brugnera:prl:107:153902}. Because of the similar pump pulse configuration, generated THz radiation may alter electron trajectories on time scales relevant to the high harmonic generation process as well [cf.\ Fig.~\ref{fig1}(c)].

\ack
Numerical simulations were performed using high performance computing resources at Rechenzentrum Garching (RZG).
We acknowledge the development of IPython~\cite{perez:cse:9:21}. ECG and SS acknowledge support by the project PRI-AIBDE-2011-0902 resp.\ DAAD-PPP-54367872. IB acknowledges the support
of DFG (project BA 41561-1).

\end{document}